\documentstyle[11pt]{article}
\newcounter{sub}
\newcounter{subeqn}[sub]
\renewcommand{\thesubeqn}{\alph{subeqn}}
\renewcommand{\theequation}{\thesub\thesubeqn}
\begin{document}

\begin{center}
{\Large{\bf  Normal modes of relativistic systems\\
in post Newtonian approximation}}\vspace{1cm}\\
Y. Sobouti$^{1,2,3}$ and V. Rezania$^1$\\
\end{center}
1. Institute for Advanced Studies in Basic Sciences, P. O. Box
 45195-159,
 Zanjan, Iran.\\
2. Physics Department, Shiraz University, Shiraz, Iran.\\
3. Center for Theoretical Physics and Mathematics, AEOI, Tehran, Iran.\\
e-mail: sobouti@sultan.iasbs.ac.ir and rezania@sultan.iasbs.ac.ir \\ \\
\vspace{3cm}\\
Publication in main journal\\
To appear under section 5, stellar cluster and associations\\
Proofs to be sent to V. Rezania
\newpage
\noindent {\bf Abstract.}
We use the post Newtonian ($pn$) order of Liouville's equation ($pnl$)
to study the normal modes of oscillation of a relativistic system.   In
addition to classical modes, we are able to isolate a new class of oscillations
that arise from perturbations of the space-time metric.    In the first
$pn$ order; a) their frequency is an order $q$ smaller than the
classical frequencies, where $q$ is a $pn$ expansion parameter;
b) they
are not damped, for there is no gravitational wave radiation in this order;
c)
they are not coupled with the classical modes in $q$ order; d) in a spherically
symmetric system, they are designated by a pair of angular momentum
eigennumbers, ($j,m$), of a pair of phase space angular momentum operators
($J^2,J_z$).    Hydrodynamical behavior of these new modes is also
investigated;
a) they do not disturb the equilibrium of the classical fluid; b) they generate
macroscopic toroidal motions that in classical case would be neutral; c) they
give rise
to an oscillatory $g_{0i}$ component of the metric tensor that otherwise is
zero for
the unperturbed system.    The conventional classical model are, of course,
perturbed to order $q$.   These, however, have not been discussed in this
paper.\\
%\end{abstract}
\section{Introduction}
In recent years some authors have debated the existence of a new class of quasi
 modes in relativistic systems believed to have been originated from the
perturbations of the space-time metric.    Kokkotas and Schutz (1986)
considered a toy model of a finite
string (to mimic a fluid) coupled to
a semi-infinite one (to substitute the dynamical space-time).
Such a
system accommodates a  family of damped oscillations.
Baumgarte and Schmidt (1993) obtained the same results by considering a
slightly more realistic toy model.     Kojima (1988) verified that strongly
damped (w-) modes do indeed exist in realistic stellar
models.      Subsequently, Kokkotas and Schutz (1992) used a WKB method to
study {\it w} mode spectra for several polytropic models.
Different investigators have
proposed different mathematical and numerical schemes to isolate these
modes.    Discrepancies exist.      Leins, Nollert and Soffel (1993)
employed two different approaches for the modes of Schwarzschild black hole.
They used Leaver's (1985) continued fraction approach and a Wronskian
technique (Nollert and Schmidt 1992).
They confirmed the findings of Kokkotas and Schutz.
They, however, found other modes with smaller oscillation frequencies
and stronger dampings than those that had been found by the WKB scheme.
Anderson, Kokkotas and Schutz (1995) used a numerical
approach to clarify the reason for the discrepancies.     They came to
agreement
with the findings of Leins et. al. (1993),and showed that the WKB
method employed by Kokkotas and Schutz failed to distinguish between these
modes.      Recently, Lindblom,
Mendell and Ipser (1997) have used a formulation of the relativistic
pulsation
equation in terms of two scalar potentials, one for the fluid and the other
for
the perturbations in the gravitational field.     They have found a set
of even-parity modes for a relativistic star, including the {\it w}
modes.

In this paper, we approach the problem through relativistic Liouville's
equation rather than relativistic hydrodynamics.    The reason for doing so, is
to avoid thermodynamic concepts incorporated into hydrodynamics.
Liouville's equation is a purely dynamical theory and free from such
complexities.     One, of course, pays the price by having to dwell in
the six
dimensional phase space an elaborate mathematical task,
but not obtrusive.      In section 2 we give the $pn$ order of the linearized
Liouville equation that governs the evolution of small perturbations from an
equilibrium distribution function.       In section 3 we extract the evolution
equation for the non classical modes and explore some of its properties.
In section 4 we explore the O(3) symmetry of the modes and classify them on
basis of this symmetries.
In section 5 we study hydrodynamics of the $pn$ modes.       In
section 6 we on a variational approach to the solution of $pn$ modes and give a
numerical analysis for the modes of polytropes.
    Lengthy mathematical manipulations
are given in appendices A and B.
\section{Formulation of the problem}
Liouville's equation in the post-Newtonian approximation $pnl$ for the
one particle distribution of a gas of collisionless particles maybe written as
follows
\stepcounter{sub}
\begin{equation}
(-i\frac{\partial}{\partial t}+ {\cal L}^{cl} +
q{\cal
L}^{^{ pn }})F({\bf x}, {\bf u}, t)=0,
\end{equation}
where $({\bf x}, {\bf u})$ are phase
space coordinates, $q$ is a small
post-Newtonian expansion parameter, the ratio of Schwarzchild radius
to a typical spatial dimension of the system, say.
The classical and post-Newtonian operators, ${\cal L}^{cl}$ and ${\cal
L}^{pn}$, respectively, are
\stepcounter{sub}
\stepcounter{subeqn}
\begin{eqnarray}
{\cal L}^{cl}&=&-i(
u^i\frac{\partial}{\partial
x^i} +\frac{\partial \theta}{\partial x^i}\frac{\partial}{\partial u^i}),\\
\stepcounter{subeqn}
{\cal L}^{^{ pn }}&=&-i[({\bf u}^2-4\theta)\frac{\partial\theta}{\partial
x^i} -u^i u^j\frac{\partial\theta}{\partial x^j}
-u^i\frac{\partial\theta}{\partial t}
+\frac{\partial\Theta}{\partial x^i}
+u^j(\frac{\partial\eta_i}{\partial x^j}-
\frac{\partial\eta_j}{\partial x^i})
+\frac{\partial\eta_i}{\partial t}]
\frac{\partial}{\partial u^i},
\end{eqnarray}
The imaginary number $i$ is included for later convenience.
The potentials
$\theta ({\bf x}, t)$, $\Theta ({\bf x}, t)$ and
$\eta\hspace{-.2cm}\eta ({\bf x}, t)$, solutions of
Einstein's equations in  $pn$  approximation, are
\stepcounter{sub}
\stepcounter{subeqn}
\stepcounter{subeqn}
\begin{eqnarray}
&&\theta({\bf x},t) = \int \frac{F({\bf x'},t,{\bf u')}}{\vert {\bf
x} - {\bf x'} \vert} d \Gamma',\;\;\;\;\;\;
\eta\hspace{-.2cm}\eta ({\bf x},t) = 4 \int \frac{{\bf
u'} F({\bf x'},t, {\bf u')}}{\vert {\bf x} - {\bf x'} \vert} d\Gamma',
\hspace{3.9cm}\;(3a,b)\nonumber\\
&&\Theta({\bf x},t) = -\frac{1}{4 \pi} \int \frac{\partial^2
F({\bf x''},t,{\bf u''})/\partial t^2 }{
\vert {\bf x} - {\bf x'} \vert \vert  {\bf
x'} - {\bf x''} \vert} d^3x'd\Gamma''
+ \frac{3}{2} \int \frac{{\bf u'}^2 F({\bf x'},t, {\bf
u'})}{\vert {\bf x} - {\bf x'} \vert} d \Gamma'\nonumber\\
\stepcounter{subeqn}
&&\hspace{2cm}- \int \frac{F({\bf x'},t,{\bf u'}) F({\bf
x''},t,{\bf u''})}{\vert {\bf x} - {\bf x'}
\vert \vert {\bf x'} - {\bf x''} \vert} d \Gamma' d
\Gamma'',
\end{eqnarray}
where $d\Gamma=d^3xd^3u$.    See Rezania and Sobouti (1998, hereafter paper
I) for details.    In an equilibrium state, $F({\bf x}, {\bf u})$ is
time-independent. If, further, it is isotropic in ${\bf u}$,
macroscopic velocities along with the vector potential
$\eta\hspace{-.2cm}\eta $ vanish.    It is also shown in
paper I that the following generalizations of the classical
energy and classical angular momentum are integrals of $pnl$:
\stepcounter{sub}
\stepcounter{subeqn}
\begin{eqnarray}
&&e=e^{cl}+qe^{^{ pn }}=\frac{1}{2}{\bf u}^2-\theta+q(2\theta^2-\Theta),\\
\stepcounter{subeqn}
&& l_i=\varepsilon_{ijk}x^ju^kexp(q\theta)\approx l_i^{cl}(1+q\theta),\;\;\;
\;for\;\;spherically\;symmetric\;\theta(r)\; and\;\Theta(r).\nonumber\\
\end{eqnarray}
Equilibrium distribution functions in  $pn$  approximation can be constructed
as appropriate functions
of these
integrals.   In paper I the $pn$ models of polytrope were studied in
this spirit.

Here we are interested in the time evolution of small deviations from
a static solution.    Let $F\rightarrow \;F(e)+ \delta F({\bf x}, {\bf u}, t)$,
$\mid\delta F\mid \ll F\;\;\;\;\forall ({\bf x}, {\bf u}, t)$.
Accordingly, the potentials split
into large and small components, $\theta(r)+\delta \theta({\bf x},
t),\;\; \Theta(r)+\delta \Theta({\bf x}, t)$ and
$\delta\eta\hspace{-.2cm}\eta({\bf x}, t)$  where
$r=\vert {\bf x}\vert$.    Both, the large and small components, can be read
out from Eqs. (3).    Substituting this splitting in Eq. (1) and keeping terms
linear in $\delta F$ gives
\stepcounter{sub}
\begin{equation}
(-i\frac{\partial }{\partial t}+ {\cal L}^{cl} +q{\cal L}^{pn})\delta F
+(\delta {\cal L}^{cl} + q\delta {\cal L}^{pn})F(e)=0,
\end{equation}
where ${\cal L}^{cl}$ and ${\cal L}^{pn}$ are now calculated from Eqs. (2) with
$\theta(r),\;\;\;\Theta(r)$ and $\eta\hspace{-.2cm}\eta=0$, and
$\delta {\cal L}^{cl}$ and $\delta {\cal L}^{pn}$ with
$\delta \theta({\bf x}, t),\;\; \delta \Theta({\bf x}, t)$ and
$\delta \eta\hspace{-.2cm}\eta ({\bf x}, t)$.    For the latters, Eqs. (2)
give
\stepcounter{sub}
\stepcounter{subeqn}
\begin{eqnarray}
&&\delta {\cal L}^{cl}F=-i\frac{\partial\delta\theta}{\partial
x^i}\frac{\partial F}{\partial u^i}=F_e{\cal L}^{cl}\delta \theta,\\
\stepcounter{subeqn}
&&\delta {\cal L}^{pn}F=F_e[{\cal L}^{cl}(\delta\Theta-4\theta\delta
\theta)+iu^2\frac{\partial\delta\theta}{\partial
t}-iu^i\frac{\partial\delta\eta_i} {\partial t}],
\end{eqnarray}
where $F_e=dF(e)/de$.    The classical limit of Eq. (5) is
\stepcounter{sub}
\begin{equation}
(-i\frac{\partial }{\partial t}+ {\cal L}^{cl})\delta F
+F_e{\cal L}^{cl}\delta\theta =0,\;\;\;\delta\theta=-G\int\frac{\delta F({\bf
x'}, {\bf u'}, t)}{\vert{\bf x}-{\bf x'}\vert}d\Gamma'.
\end{equation}
Equation (7) is the linearized Liouville-Poisson equation.    It was studied
briefly by Antonov (1962).     He separated $\delta F$
and Eq. (7) into even and odd components in ${\bf u}$, extracted the equation
governing the odd component and showed that $\delta F_{odd}$ satisfies an
eigenvalue equation.    Sobouti (1984, 1985, 1986, 1989a, b)
elaborated on
this eigenvalue problem, studied some of its symmetries and approaches to its
solution.    Sobouti and Samimi (1989), and Samimi and Sobouti (1995) showed
that Eq. (7) has an O(3) symmetry and its oscillation modes
can be classified by the eigennumber $j$ and $m$ of certain phase space angular
momentum operator.

The effect of the  $pn$  corrections in Eq. (5) on the known solutions of the
classical Eq. (7) can be analyzed by the usual perturbation techniques.
Whatever the procedure, the first order corrections on the known oscillation
modes will be small and will not change their nature.    We will not pursue
such issues here.    The main interest of the this paper is to study a new
class of solutions of Eq. (5) that originate solely from the  $pn$  terms
and have no precedence in classical theories.    It is not difficult to
anticipated the existence of such modes.     Functions of the classical
energy and angular momentum satisfy Eq. (7) with no time dependence.
They may be considered
as the eigenfunctions of that equation belonging to the infinitely degenerate
eigenvalue $\omega =0$.    Now the  $pn$  corrections of Eq. (5) set up an
eigenvalue equation in this degenerate subspace and give rise to a new class of
modes that have relativistic origin and are not coupled to the classical
solutions, in the first  $pn$  order.
\section{The post-Newtonian modes of oscillations}
{\it Theorem:} Any function $G(e^{cl}, {\bf l}^{cl})$ of classical energy and
angular momentum, $e^{cl}$ and ${\bf l}^{cl}$, respectively, for which
$\delta\rho ({\bf x})=\int G d^3u=0$, is a static solution of the linearized
Liouville-Poisson, Eq. (7).    For, ${\cal L}^{cl}G=0$ and $\delta\theta ({\bf
x})=\int\delta\rho ({\bf x'})\vert {\bf x}-{\bf x'}\vert^{-1}d^3x'=0$, QED.

One class of such functions are those with odd parity in ${\bf u}$ or,
equivalently, in ${\bf l}$, i. e., $G(e^{cl}, {\bf l}^{cl})= -G(e^{cl},- {\bf
l}^{cl})$.        We will see in Eqs. (17) and (18) below that
there exists a class of odd parity functions that not only fulfill the
requirements of the theorem but also give $\delta\Theta =0$.     With these
preliminaries, we come back to solutions of Eq. (5).

Let $\delta F({\bf x}, {\bf u}, t)=\delta F(e^{cl}, l_i^{cl})e^{-i\omega t}$,
where $\delta F(e^{cl},\;l_i^{cl})$ is a static solution of Eq. (7).    By the
theorem above, Eqs. (5) and (6) reduce to
\stepcounter{sub}
\begin{equation}
q{\cal L}^{pn}\delta F =\omega(\delta F+ qF_e u^i\delta\eta_i).
\end{equation}
Furthermore the term containing $\delta\eta\hspace{-.2cm}\eta$ is an order
$q$ smaller than $\delta F$ and can be dropped.   Thus
\stepcounter{sub}
\begin{equation}
q{\cal L}^{pn}\delta F=\omega\delta F.
\end{equation}
where
\stepcounter{subeqn}
\begin{equation}
{\cal L}^{pn}=\frac{-i}{r}\{[(u^2-4\theta)\theta'+\Theta']x^i-\theta'({\bf
x}\cdot {\bf u})u^i\}\frac{\partial}{\partial u^i},\;\;``\;'\;" =\frac{d}{dr},
\end{equation}
see Eq. (A. 1) for details.    We owe the reader a
demonstration of $\delta\Theta=0$.
In appendix A we show that $\omega$ is real.    There follows that $\delta F$
is complex; for, ${\cal L}^{pn}$ is purely imaginary.    Thus, $\delta F=
\delta F_{_R} + i\delta F_{_I}$.      Equation (9) can be decomposed
accordingly:
\stepcounter{sub}
\stepcounter{subeqn}
\begin{eqnarray}
&&q{\cal L}^{pn}\delta F_{_R}=i\omega\delta F_{_I} ,\\
\stepcounter{subeqn}
&&q{\cal L}^{pn}\delta F_{_I}=-i\omega\delta F_{_R}.
\end{eqnarray}
Upon eliminating the imaginary component one obtains
\stepcounter{sub}
\begin{equation}
q^2{{\cal L}^{pn}}^2\delta F_{_R}=\omega^2\delta F_{_R}.
\end{equation}
The same equation is satisfied by $\delta F_{_I}$.     The latter will,
however, be obtained from Eq. (10a) once $\delta F_{_R}$ is calculated from Eq.
(11).    In appendix A we show that ${{\cal L}^{pn}}^2$ is Hermitian, in
spite of the fact that ${\cal L}^{pn}$ is not.  Furthermore,  ${{\cal
L}^{pn}}^2$ is  even in both ${\bf
x}$ and ${\bf u}$.    Therefore, the eigensolutions $\delta F_{_R}$ can be
chosen either odd or even in ${\bf x}$ and ${\bf u}$.    Even parities,
however, are not
acceptable.    For they give rise to non vanishing changes in the macroscopic
density, $\delta\rho$, and the potential $\delta\theta$ and $\delta\Theta$,
which violates the validity of Eq. (8).     The $x$ and $u$-parities of
$\delta F_{_I}$, however, will be opposite to those of $\delta F_{_R}$; for they
are
connected by Eqs. (10a, b), where ${\cal L}^{pn}$ is odd in ${\bf x}$ and
${\bf u}$.
\section{O(3) symmetry of ${\cal L}^{pn}$}
In appendix A we show that ${\cal L}^{pn}$ and ${{\cal L}^{pn}}^2$
commute with the phase space angular momentum operators
$J_i=-i\varepsilon_{ijk}(x^j\frac{\partial}{\partial x^k}+
u^j\frac{\partial}{\partial u^k})$.    Thus, they can have simultaneous
eigensolutions with $J^2$ and $J_z$.    Eigensolutions of the latter pair of
operators,
$\Lambda_{jm}$; $j$ and $m$ integers, are worked out in appendix B.    They are
complex polynomials of the components of the classical angular
momentum vector.    Their $x$ and $u$  parity is that of $j$.

An expression of the form $f(e^{cl},\;{l^{cl}}^2)\;\Lambda_{jm}$ is already
an eigensolution of $J^2$ and $J_m$ belonging to eigennumbers $j$ and $m$.  See
appendix B, Eqs. (B. 2-5).      Solutions of Eq. (11) for $\delta F_R$ will,
therefore, reduce to substituting an expression of this sort with an odd $j$
value in Eq. (11) and solving it for $f(e^{cl},\;{l^{cl}}^2)$.        In
section 6 we will take a varitional approach to such solutions, and as
variational trial functions will consider functions of $e^{cl}$ only.   For the
moment let
\stepcounter{sub}
\begin{equation}
\delta F_R=f_{jm}=f(e)Re\;\Lambda_{jm}=[\sum_{n=j+1}^N c_n (-e)^n] Re
\Lambda_{jm},\;\;\;\;\;\;\;c_n=consts,
\end{equation}
be a solution of Eq. (11) belonging to the eigenvalue $\omega$.     Combining
this with its corresponding imaginary component from Eq. (10a) we obtain
\stepcounter{sub}
\begin{equation}
\delta F_{jm}({\bf x}, {\bf u}, t)=(1+\frac{q}{\omega}{\cal L}^{pn})f_{jm}e^{-i
\omega t}.
\end{equation}
This expression satisfies Eq. (9), if $f_{jm}$ satisfies Eq. (11).   At this
stage let us note an important property of Liouville's equation.    If a pair
$(\omega, \delta F)$ is an eigensolution of Liouville's equation,
$(-\omega, \delta F^*)$ is another eigensolution.   This can be verified by
taking the complex conjugate of Eq. (9).    These solutions, being complex
quantities, cannot serve as physically meaningful distribution functions.
Their real or imaginary part, including time dependence, however, can.    With
no lost of generality we will adopt the real part.    Thus,
\stepcounter{sub}
\begin{equation}
Re\;\delta F_{jm}({\bf x}, {\bf u},
t)=f(e)Re\;\Lambda_{jm}\cos\omega t +i\frac{q}{\omega}{\cal L}^{pn}(f(e)
Re\;\Lambda_{jm})\sin\omega t.
\end{equation}
This, however, is the sum of two eigenfunctions belonging to $\omega$ and
$-\omega$ and is not a simple eigensolutions of Eq. (9).
\section{Hydrodynamics of $pn$ modes}
In this section we calculate the density fluctuations, macroscopic velocities,
and the perturbations in the space-time metric generated by a $pn$ mode.

For $j$ an odd integer, $f_{jm}({\bf x}, {\bf u})$ is odd while ${\cal
L}^{pn}f_{jm}$ is even in ${\bf x}$ and ${\bf u}$.
The macroscopic velocities are obtained by multiplying Eq. (14) by ${\bf u}$
and integrating over the u-space.     Only the odd component of $\delta F_{jm}$
contributes to this bulk motion,
\stepcounter{sub}
\begin{equation}
\rho {\bf v}=\int f(e) Re\;\Lambda_{jm}{\bf u} d^3u\; \cos\omega t.
\end{equation}
In appendix B we show that $\rho {\bf v}$ is a toroidal spherical harmonic
vector field.
In spherical polar coordinates it has the following form
\stepcounter{sub}
\stepcounter{subeqn}
\begin{equation}
\rho (v_r,\; v_{\vartheta},\; v_{\varphi})=r^jG(v_{es})(0,\;
Re\;\frac{-1}{\sin\vartheta}\frac{\partial}{\partial\varphi}Y_{jm}(\vartheta,
\varphi),\; Re\;
\frac{\partial Y_{jm}}{\partial\vartheta}(\vartheta, \varphi))\;\cos\omega t,
\end{equation}
where
\stepcounter{subeqn}
\begin{equation}
G(v_{es})=\int_0^{v_{es}}f(e)u^{j+3}du,
\end{equation}
and $v_{es}=\sqrt{2\theta}$ is the escape velocity from the potential $\theta
(r)$.
The macroscopic density,
generated by even component of Eq. (14), is
\stepcounter{sub}
\begin{eqnarray}
&&\delta \rho ({\bf x}, t)=i\frac{q}{\omega}\int
{\cal L}^{pn}(f(e)Re\;\Lambda_{jm})d^3u \sin\omega t\nonumber\\
&&\hspace{1.4cm}=2\frac{q}{\omega}\frac{\theta'}{r}{\bf x}\cdot\int
f(e)Re\;\Lambda_{jm}{\bf u}d^3u\; \sin\omega t =0.
\end{eqnarray}
The second integral is obtained by an integration by parts.    The vanishing of
it comes about because of the fact that the radial vector
${\bf x}$ is orthogonal to the toroidal vector $\rho {\bf v}$.
One also notes that $\nabla\cdot (\rho {\bf v})=0$.
It can further be verified taht,  the continuity
equation is satisfied at both classical and $pn$ level.
To complete the proof of validity of Eq. (9) one should also show that
$\delta\Theta=0$.     From the definition Eq. (3c) for $\delta\Theta$ and Eq.
(14) for $\delta F$ one has
\stepcounter{sub}
\begin{eqnarray}
&&\delta\Theta = \frac{\omega^2}{4\pi} \int \frac{\delta\theta
({\bf x'})}{
\vert {\bf x} - {\bf x'} \vert} d^3x'
+ \int\frac{\rho (r')\delta\rho ({\bf x''})+\delta\rho({\bf
x'})\rho (r'')}{\vert {\bf x} - {\bf x'}
\vert \vert {\bf x'} - {\bf x''} \vert} d^3x' d^3x''\nonumber\\
&&\hspace{2cm}+ \frac{3}{2}\int\frac{d^3x'}{\vert {\bf x} - {\bf x'} \vert}\int
{\bf u'}^2 \delta F({\bf x'}, {\bf u'}) d^3u'.
\end{eqnarray}
The first two terms are zero because $\delta\rho$ and $\delta\theta$ are zero.
The third term vanishes because the integral over ${\bf u'}$ has the same form
as in $\delta\rho$ except for the additional scalar factor ${\bf u'}^2$ in the
integrand.    Like $\delta\rho$ it can be reduce to the inner product of
the radial
vector ${\bf x}$ and a toroidal vector that are orthogonal to each other. QED.

The toroidal motion
described here
slides one spherical shell of the fluid over
the other without perturbing the density, the Newtonian gravitational field
and,
therefore, the hydrostatic equilibrium of the classical fluid.    In doing so,
it does not
affect and is not affected by the conventional classical modes of the fluid at
this first $pn$ order.

Nonetheless, the $pn$ modes are associated with space time perturbations.
From Eq. (8c) of paper I and Eq. (3b) of this paper, $g_{0i}$ component of the
metric tensor is
\stepcounter{sub}
\begin{equation}
g_{0i}=\eta_i=4\int\frac{\rho v_i({\bf x'})}{\vert {\bf x}-{\bf x'}\vert}d^3x'.
\end{equation}
In spherical polar coordinates, one has
\stepcounter{sub}
\stepcounter{subeqn}
\begin{eqnarray}
&&\eta_r=0,\\
\stepcounter{subeqn}
&&\eta_{\vartheta}=-a_jRe\;\frac{1}{\sin\vartheta}
\frac{\partial}{\partial\varphi}Y_{jm}(\vartheta,\varphi)\cos\omega t,\\
\stepcounter{subeqn}
&&\eta_{\varphi}=a_j
Re\;\frac{\partial Y_{jm}}{\partial\vartheta}(\vartheta,\varphi)\cos\omega t,
\end{eqnarray}
where
\stepcounter{subeqn}
\begin{eqnarray}
&&a_j=\frac{16\pi}{2j+1}\left\{ \begin{array}{ll}
                               (r/R)^jy_j(R)+(2j+1)r^j\int_r^R{r'}^{-j-1}y_j(r')dr'&\;\;for\;r<R\\
                               (R/r)^{j+1}y_j(R)&\;\;for\;r>R
                                 \end{array}\right.           \\
\stepcounter{subeqn}
&&y_j(r)=r^{-j-1}\int_0^r{r'}^{2j+2}G(\theta(r'))dr',\\
\stepcounter{subeqn}
&&G(\theta (r))=\int_0^{v_{es}}f(e)u^{j+3}du\nonumber  \\
&&\hspace{1.5cm} =2^{j/2+1}\Gamma(j/2+2)
\Gamma(n+1) \theta(r)^{n+j/2+2}/\Gamma(n+j/2+3),
\end{eqnarray}
where $R$ is the radius of the system and $\Gamma(n)$ is the gamma function.
The remaining components of the metric tensor remain unperturbed.
\section{Variational solutions of $pn$ modes}
In Eq. (12) we suppress the subscript $jm$, for
brevity.    It will be remembered that $f$ will be chosen odd in both ${\bf x}$
and ${\bf
u}$ and will be a function of the phase coordinates through the classical
energy and
angular momentum integrals.    We multiply Eq. (11) by $f^*$ and integrate
over the volume of the phase space available to the system
\stepcounter{sub}
\stepcounter{subeqn}
\begin{equation}
\omega^2\int f^2 d\Gamma=q^2\int f{{\cal
L}^{pn}}^2 f d\Gamma
=q^2\int({{\cal L}^{pn}}^{\dagger}f)^*
({\cal L}^{pn} f)d\Gamma.
\end{equation}
The adjoint operator ${{\cal L}^{pn}}^{\dagger}$ is given in Eq. (A. 5).
Substituting the latter in Eq. (21a) and simple manipulations gives
\stepcounter{subeqn}
\begin{equation}
\frac{\omega^2}{q^2}\int f^2 d\Gamma=\int\vert{\cal
L}^{pn} f \vert^2 d\Gamma
+\int\theta'^2[u^2-3\frac{({\bf x}\cdot{\bf u})^2}{r^2}]
f^2 d\Gamma.
\end{equation}
It will be remembered that $\omega$ is real and $\omega^2$ is positive.
Therefore, in spite of one negative term, the right hand side of Eq. (21b) has
to be positive.   That is, ${{\cal L}^{pn}}^2$ is positive in addition to being
Hermitian.
To solve Eq. (21b) by
variational techniques we choose the set of trial functions
$\{(-e)^n\Lambda_{jm},\; j=odd,\;n>j,\; e=u^2/2-\theta\}$.    Any member of this
set is an eigenfunction of $J^2$ and $J_z$ with eigenvalues $j$ and $m$,
respectively.    Next we choose a real linear combination of these trial
functions as an ansatz for $f$
\stepcounter{sub}
\begin{equation}
f_{jm}=f(e)Re\;\Lambda_{jm}=[\sum_{n=j+1}^N c_n (-e)^n] Re
\Lambda_{jm},\;\;\;\;\;\;\;c_n=consts.
\end{equation}
We substitute Eq. (22) in (21b), carry out the integrations, and minimize
$\omega^2$ with respect to variations of $c_n$'s.     We arrive at the
following matrix equation to solve for $\omega^2$ and $c_n$'s:
\stepcounter{sub}
\begin{equation}
WC=\frac{\omega^2}{q^2}SC,
\end{equation}
where $C$ is the column vector  of the variational parameters, $\{c_n\}$, and
elements of the $S$ and $W$ matrices are
\stepcounter{sub}
\stepcounter{subeqn}
\begin{eqnarray}
&&S_{pq}=\int(-e)^{p+q}\vert Re\;\Lambda_{jm}\vert^2 d\Gamma,\\
\stepcounter{subeqn}
&&W_{pq}=\int({\cal L}^{pn}(-e)^p Re\;\Lambda_{jm})^*
({\cal L}^{pn}(-e)^q Re\;\Lambda_{jm})
+\int\theta'^2[u^2-3\frac{({\bf x}\cdot{\bf u})^2}{r^2}]
(-e)^{p+q}\vert Re\;\Lambda_{jm}\vert^2 d\Gamma.\nonumber\\
\end{eqnarray}
Eigen $\omega$'s are the roots of the characteristic equation
\stepcounter{sub}
\begin{equation}
\vert W-\frac{\omega^2}{q^2}S\vert=0.
\end{equation}
For each $\omega$, Eq. (23) can then be solved for the eigenvector C.
This completes the Rayleigh-Ritz variational formalism of solving Eq. (11).
In what follows we present some numerical values for polytropes.
\vspace{3cm}\\
{\large{\bf pn Modes of polytropes belonging to $(j, m)=(1, 0)$}}\\
From Eq. (B. 9),
$\Lambda_{1\;0}=l_z=ru\cos\gamma$, where $\gamma$ is the angle between ${\bf
x}$ and ${\bf u}$.    Substituting this in Eqs. (24) and integrating over
directions of ${\bf x}$ and ${\bf u}$ vectors and over $0<u<\sqrt{2\theta}$
gives
\stepcounter{sub}
\stepcounter{subeqn}
\begin{eqnarray}
&&S_{pq}=\int_0^1\theta^{p+q+2.5}x^4dx,\\
\stepcounter{subeqn}
&&W_{pq}=4\pi
G\rho_c\{(16a_{pq}-b_{pq})\int_0^1{\theta'}^2\theta^{p+q+3.5}x^4dx \nonumber\\
&&\hspace{1.7cm}+(1-8a_{pq})\int_0^1\Theta'\theta'\theta^{p+q+2.5}x^4dx
\nonumber\\
&&\hspace{2.8cm}+a_{pq}\int_0^1{\Theta'}^2\theta^{p+q+1.5}x^4dx\},\\
\stepcounter{subeqn}
&&a_{pq}=\frac{pq(p+q+2.5)}{(p+q)(p+q-1)},\;\;\;b_{pq}= \frac{4p+4q+ 11}{p+q+
3.5},\;\;\;p,\;q\;=2, 3, \cdots.
\end{eqnarray}
Polytropic potentials $\theta$ and $\Theta$ were obtained from integrations of
Lane Emden equation and Eqs. (28) of paper I, respectively.    Eventually,
the matrix elements of Eqs. (26), the characteristic Eq. (25) and the
eigenvalue Eq. (23) were numerically solved in succession.   Tables 1-4 show
some sample calculations for polytropes 2, 3, 4, and 4.9.     Eigenvalues are
displayed in lines marked by an asterisks.     The column following an
eigenvalue is the corresponding eigenvector, i. e. the values of $c_1,\;c_2,\;
\cdots$, of Eq. (22).    To demonstrate the accuracy of the procedure,
calculations with six and seven variational parameter are given for comparison.
The first three eigenvalues can be trusted up to four to two figures.
Convergence improve as the polytropic index, i. e. the central condensation,
increases.    Eigenvalues are in units of $4\pi G\rho_c q^2$ and increase as
the mode order increases.

\newpage
\setcounter{sub}{0}
\setcounter{subeqn}{0}
\renewcommand{\theequation}{A.\thesub\thesubeqn}
\noindent {\large{\bf Appendix A: Some properties of ${\cal L}^{pn}$ and
${{\cal L}^{pn}}^2$}}\\
{\it The Hilbert space:} let ${\cal H}$ be the space of complex square
integrable functions of phase coordinates $({\bf x}, {\bf u})$ that
vanish at the phase space boundary of the system,
$${\cal H}:\{g({\bf x}, {\bf u}),\;\int g^*gd\Gamma=finite,\;g(boundary)=0\}.$$
Integrations in ${\cal H}$ are over the volume of phase space available to the
system.   In particular, the boundedness of the system in velocity space
sets
the upper limit of $u$ at the escape velocity, $\sqrt{2\theta}$, where $\theta
({\bf x})$ is the potential at ${\bf x}$.    Thus, $g({\bf x}, \sqrt{2\theta
({\bf x})})=0$.

The operator ${\cal L}^{pn}$ is defined in ${\cal H}$.    For a spherically
symmetric, isotropic and static system, Eq. (2b) gives
\stepcounter{sub}
\begin{equation}
{\cal L}^{pn}=\frac{-i}{r}\{[(u^2-4\theta)\theta'+\Theta']x^i-\theta'({\bf
x}\cdot {\bf u})u^i\}\frac{\partial}{\partial u^i}.
\end{equation}
We have set $\eta\hspace{-.2cm}\eta$ and all time
derivatives
equal to zero, used $\partial /\partial x_i=(x_i/r)d/dr$ and denoted $d/dr$ by
``$\prime$ ''.\\
{\it Theorem 1.} Eigenvalues of ${\cal L}^{pn}$ are real:    This is in spite
of the fact that ${\cal L}^{pn}$ is not Hermitian. \\  {\it Proof:} Let
\stepcounter{sub}
\begin{equation}
{\cal L}^{pn}f=\omega f,\;\;\;\;\; f\in {\cal H}.
\end{equation}
Noting that ${\cal L}^{pn}$ is purely imaginary, the complex conjugate of this
equation is
\stepcounter{sub}
\begin{equation}
{\cal L}^{pn}f^*=-\omega^* f^*.
\end{equation}
Multiplying (A. 2) by $f^*$ and (A. 3) by $f$, adding the two equations and
integrations over the phase space gives
\stepcounter{sub}
\begin{eqnarray}
&&0=\int(f^*{\cal L}^{pn}f+ f{\cal L}^{pn}f^*)d\Gamma=\int {\cal
L}^{pn}(f^*f)d\Gamma\nonumber\\
&&\hspace{4.3cm}=(-2i)\int\theta'\frac{{\bf x}\cdot{\bf
u}}{r}f^*fd\Gamma,\nonumber\\
&&\hspace{4.3cm}=(\omega-\omega^*)\int f^*fd\Gamma.
\end{eqnarray}
The ${\bf x}$ and ${\bf u}$ parities of $f^*f$ are even
and those of ${\bf x}\cdot{\bf u}$ are odd.
Therefore, the integral vanishes and $\omega=\omega^*=real$. QED.\\
{\it Hermitian adjoint of ${\cal L}^{pn}$:}  For $f,\; g\in {\cal H}$,
by an integration by parts one finds:
$$\int g^*{\cal L}^{pn}f d\Gamma=\int[({\cal L}^{pn} + 2i\frac{{\bf x}\cdot{\bf
u}}{r}\theta')g]^*fd\Gamma,$$
thus
\stepcounter{sub}
\begin{equation}
{{\cal L}^{pn}}^{\dagger}={\cal L}^{pn}+2i\frac{{\bf x}\cdot{\bf u}}{r}\theta'.
\end{equation}
{\it Theorem 2.   ${{\cal L}^{pn}}^2$ is Hermitian in ${\cal H}$}.
{\it Proof:}   For $g\in {\cal H}$
\stepcounter{sub}
\begin{eqnarray}
&&\int g^*{{\cal L}^{pn}}^2 gd\Gamma=\int ({{\cal L}^{pn}}^{\dagger}g)^*
{\cal L}^{pn} g d\Gamma\nonumber\\
&&\hspace{2cm}=\int [({\cal L}^{pn}+2i\frac{{\bf x}\cdot{\bf u}}{r}\theta'g]^*
{\cal L}^{pn}g f\Gamma\nonumber\\
&&\hspace{2cm}=\int\vert {\cal L}^{pn}g\vert ^2d\Gamma+ \int
{\theta'}^2 [u^2-3\frac{({\bf x}\cdot{\bf u})^2}{r^2}]g^* gd\Gamma=real,
\end{eqnarray}
QED.   Some integrations by parts are carried out in calculating the
term containing ${\bf x}\cdot{\bf u}$.\\
{\it O(3) symmetry of ${\cal L}^{pn}$ and ${{\cal L}^{pn}}^2$:}    For a
spherically symmetric system, ${\cal L}^{pn}$ of Eq. (A. 1) depends on the
angle
between ${\bf x}$ and ${\bf u}$ and their magnitudes.   Simultaneous rotations
of the $x$ and $u$ coordinates about the same axis by the same
angle should leave ${\cal L}^{pn}$ form invariant.    The generator of such
simultaneous infinitesimal rotations on the function space ${\cal H}$ is
\stepcounter{sub}
\begin{equation}
J_i=J_i^{\dagger}=-i\varepsilon_{ijk}(x^j\frac{\partial}{\partial x^k}
+u^j\frac{\partial}{\partial u^k}).
\end{equation}
With $\theta$ and $\Theta$ of Eq. (A. 1) being spherically symmetric, it is
easy to show that ${\cal L}^{pn}$ commutes with $J_i$
\stepcounter{sub}
\begin{equation}
[{\cal L}^{pn},J_i]=0.
\end{equation}
The operators $J_i$, however, obey the angular momentum algebra.    Therefor,
the three operators ${\cal L}^{pn},\;J^2$ and $J_z$ commute pairwise.     The
same is true of ${{\cal L}^{pn}}^2,\;J^2$ and $J_z$.    In appendix B we
elaborate on the eigenfunctions of $J^2$ and $J_z$ to prepare the path for
simultaneous eigensolutions of the latter trio.\\

\newpage
\setcounter{sub}{0}
\setcounter{subeqn}{0}
\renewcommand{\theequation}{B.\thesub\thesubeqn}
\noindent {\large{\bf Appendix B: Eigensolutions of  $J^2$ and $J_z$}}\\
As pointed out earlier, $J_i$'s of Eq. (A. 7) have the angular momentum
algebra,
\stepcounter{sub}
\begin{equation}
[J_i,J_j]=i\varepsilon_{ijk}J_k.
\end{equation}
The simultaneous eigensolutions of $J^2$ and $J_z$, $\Lambda_{jm}({\bf x}, {\bf
u})$, obey the followings
\stepcounter{sub}
\begin{equation}
J^2\Lambda_{jm}=j(j+1)\Lambda_{jm},\;\;\;\;\;j=0,1,\cdots,
\end{equation}
\stepcounter{sub}
\begin{equation}
J_z\Lambda_{jm}=m\Lambda_{jm},\;\;\;\;\;-j\le m\le j.
\end{equation}
The ladder operators, $J_{\pm}=J_x\pm iJ_y$, raise and lower $m$ values:
\stepcounter{sub}
\begin{equation}
J_{\pm}\Lambda_{jm}=\sqrt{(j\mp m)(j\pm
m+1)}\Lambda_{j m\pm 1}.
\end{equation}
In particular
\stepcounter{subeqn}
\begin{equation}
J_{\pm}\Lambda_{j,\pm j}=0.
\end{equation}
The effect of $J_i$ on classical energy integral, $e=u^2/2-\theta (r)$, and
the classical angular momentum integral, $l_i=\varepsilon_{ijk}x_ju_k$, are as
follows
\stepcounter{sub}
\stepcounter{subeqn}
\begin{eqnarray}
&&J_i e=J_i l^2=J_i f(e, l^2)=0,\\
\stepcounter{subeqn}
&&J_i l_j=i\varepsilon_{ijk} l_k.
\end{eqnarray}
It is simple to verify Eqs. (B. 5) by direct substitution.\\
{\it Theorem 1:}
\stepcounter{sub}
\begin{equation}
\Lambda_{j,\pm j}=l_{\pm}^j=(\frac{1}{2})^j(l_x\pm il_y)^j.
\end{equation}
{\it Proof:}
\stepcounter{sub}
\stepcounter{subeqn}
\begin{eqnarray}
&&J_zl_{\pm}^j=jl_{\pm}^{j-1}(J_zl_{\pm})=\pm
jl_{\pm}^j,\hspace{2.6cm}by\; (B. 5b),\\
\stepcounter{subeqn}
&&J^2l_+^j=(J_-J_++J_z^2+J_z)l_+^j=j(j+1)l_+^j,\;\;\;by\;
(B. 4a)\;and\;(B. 7a),\\
\stepcounter{subeqn}
&&J^2l_-^j=(J_+J_-+J_z^2-J_z)l_-^j=j(j+1)l_-^j,
\end{eqnarray}
QED.   Combining Eqs. (B. 6), (B. 4) and (B. 5) one obtains
\stepcounter{sub}
\begin{equation}
\Lambda_{jm}=af(e, l^2)J_+^{j-m}l_-^j=bf(e, l^2)J_-^{j-m}l_+^j,
\end{equation}
where $f(e, l^2)$ is an arbitrary function of its arguments, and $a$ and $b$
are normalization constants.    Examples:
\stepcounter{sub}
\stepcounter{subeqn}
\begin{eqnarray}
&&\Lambda_{1\;0}=l_z,\\
\stepcounter{subeqn}
&&\Lambda_{1\;\pm1}=l_{\pm},\\
\stepcounter{subeqn}
&&\Lambda_{2\;0}=2l_+l_--l_z^2=\frac{1}{2}(3l_z^2-l^2),\\
\stepcounter{subeqn}
&&\Lambda_{2\;\pm 1}=l_{\pm}l_z,\\
\stepcounter{subeqn}
&&\Lambda_{2\;\pm 2}=l_{\pm}^2.
\end{eqnarray}
{\it Theorem 2:}   The vector field ${\bf V}^{jm}=\int\Lambda_{jm}{\bf
u}d\Omega$ is a toroidal vector field belonging to the spherical harmonic
numbers ($j,m$), where integration is over the directions of ${\bf u}$.\\
{\it Preliminaries:}   Let ($\vartheta,\varphi$) and ($\alpha,\beta$), and
$\gamma$ denote the polar angles of ${\bf x}$, of ${\bf u}$ and the angles
between (${\bf x}, {\bf u}$), respectively.     Also choose magnitudes of ${\bf
x}$ and ${\bf u}$ to be unity, as only integrations over the
direction angles are of concern.     One has $\cos\gamma
=\cos\vartheta\;\cos\alpha\;+\;\sin\vartheta\;\sin\alpha\;\cos(\varphi-\beta)$
\stepcounter{sub}
\stepcounter{subeqn}
\begin{eqnarray}
&&u_r=\cos\gamma ,\\
\stepcounter{subeqn}
&&u_{\vartheta}=-\sin\vartheta\;\cos\alpha\;+\;\cos\vartheta\;\sin\alpha\;
\cos(\varphi-\beta),\\
\stepcounter{subeqn}
&&u_{\varphi}=-\sin\alpha\;\sin(\varphi-\beta),\\
\stepcounter{subeqn}
&&l_+=i(\sin\vartheta\;\cos\alpha\;e^{i\varphi}-\cos\vartheta\;\sin\alpha\;e^{
 i\beta}).
\end{eqnarray}
{\it Proof:}   By induction, we show that a) ${\bf V}^{jj}$ is a toroidal field
and b) if ${\bf V}^{jm}$ is a toroidal field, so is ${\bf V}^{j\;m-1}$.\\
a) Direct integrations over $\alpha$ and $\beta$ gives
\stepcounter{sub}
\stepcounter{subeqn}
\begin{eqnarray}
&&V_r^{jj}=\int l_+^j u_r
d\Omega=0,\;\;\;\;\;\;d\Omega =\sin\alpha\;d\alpha\;d\beta,\\
\stepcounter{subeqn}
&&V_{\vartheta}^{jj}=\int l_+^j u_{\vartheta}
d\Omega=-\frac{1}{\sin\vartheta}\frac{\partial}{\partial\varphi}Y_{jj}(
\vartheta, \varphi), \\
\stepcounter{subeqn}
&&V_{\varphi}^{jj}=\int l_+^j u_{\varphi}
d\Omega=\frac{\partial}{\partial\vartheta}Y_{jj}(\vartheta,
\varphi).\;\;QED.
\end{eqnarray}
b)  Suppose ${\bf V}^{jm}$ is a toroidal vector field and calculate ${\bf
V}^{j\;m-1}=
\int(J_-\Lambda_{jm}){\bf u}d\Omega$, where $J_{\pm}=L_{\pm}+K{\pm}$,
$L_{\pm}=
\pm e^{\pm i\varphi}(\frac{\partial}{\partial\vartheta}\pm i cotg\vartheta
\frac{\partial}{\partial\varphi})$,
$K_{\pm}=
\pm e^{\pm i\beta}(\frac{\partial}{\partial\alpha}\pm i cotg\alpha
\frac{\partial}{\partial\beta})$.     Again  direct integrations gives
\stepcounter{sub}
\stepcounter{subeqn}
\begin{eqnarray}
&&V_r^{j\;m-1}=L_-V_r^{jm}=0,\hspace{3cm} if\;\; V_r^{jm}=0,\\
\stepcounter{subeqn}
&&V_{\vartheta}^{j\;m-1}=
-\frac{1}{\sin\vartheta}\frac{\partial}{\partial\varphi}Y_{j\;m-1}(
\vartheta, \varphi), \hspace{1cm} if\;\; V_{\vartheta}^{j\;m}=
-\frac{1}{\sin\vartheta}\frac{\partial}{\partial\varphi}Y_{j\;m}(
\vartheta, \varphi),\\
\stepcounter{subeqn}
&&V_{\varphi}^{j\;m-1}=
\frac{\partial}{\partial\vartheta}Y_{j\;m-1}(\vartheta,
\varphi),\hspace{1cm} if\;\; V_{\varphi}^{j\;m}=
\frac{\partial}{\partial\vartheta}Y_{j\;m}(
\vartheta, \varphi).
\end{eqnarray}
QED.
\newpage
\noindent {\large{\bf References}\vspace{1cm}\\
Andersson, N., Kokkotas, K. D., Schutz, B. F., 1995, M. N. R. A. S.,
{\bf 274}, 1039\\
Antonov, V. A., 1962, Vestnik Leningerad gos. Univ.,
{\bf 19}, 69\\
Baumgarte, T. W., Schmidt, B. G., 1993, Class. Quantum. Grav.,
{\bf 10}, 2067\\
Kojima, Y., 1988, Prog. Theor. Phys.,
{\bf 79}, 665\\
Kokkotas, K. D., Schutz, B. F., 1986, Gen. Relativ. Gravitation,
{\bf 18}, 913\\
Kokkotas, K. D., Schutz, B. F., 1992, M. N. R. A. S.,
{\bf 254}, 119\\
Leaver, E. W., 1985, Proc. R. Soc. London A,
{\bf 402}, 285\\
Leins, M., Nollert, H. P., Soffel, M. H., 1993, Phys. Rev. D,
{\bf 48}, 3467\\
Lindblom, L., Mendell, G., Ipser, J. R., 1997, Phys. Rev. D,
{\bf 56}, 2118\\
Nollert, H. -P., Schmidt, B. G., 1992, Phys. Rev. D,
{\bf 45}, 2617\\
Samimi, J., Sobouti, Y., 1995, A \& A, {\bf 297}, 707\\
Sobouti, Y., 1984, A \& A, {\bf 140}, 821;
1985, {\bf 147}, 61;
1986, {\bf 169}, 95;
1989a, {\bf 210}, 18;
1989b, {\bf 214}, 83\\
Sobouti, Y., Samimi, J., 1989, A \& A, {\bf 214}, 92\\
\newpage
\noindent Table 1: $pn$ mods of polytrope n=2, belonging to $(j, m)=(1,0)$.
Eigenvalues are in units $4\pi G\rho_c q^2$.    $c_n$'s are linear variational
parameters.   The first three eigenvalues are reliable up to three figures.
Higher order eigenvalues are less accurate.   A number $a\times 10^{\pm b}$ is
written a $a\pm b$.
\vspace{.05cm}\\
\noindent Table 2: Same as Table 1. $n=3$ and $(j,m)=(1,0)$.
\vspace{.05cm}\\
\noindent Table 3: Same as Table 1. $n=4$ and $(j,m)=(1,0)$.
\vspace{.05cm}\\
\noindent Table 4: Same as Table 1. $n=4.9$ and $(j,m)=(1,0)$.
\vspace{.05cm}\\
\newpage
\begin{center}
Table 1.\vspace{.2cm}\\
\begin{tabular}{crrrrrrr}\hline
$\omega^2$& .1825+01& .4973+01& .6448+01& .1216+02& .3425+02& .1686+03&\\
&&&&&&&\\
$c_1$& .3113+02&-.8912+02& .1663+03& .1344+03& .7545+01&-.1399+04&\\
$c_2$& .3908+02& .1045+04&-.3234+04&-.9746+03&-.2392+04& .8484+04&\\
$c_3$&-.1420+03&-.6649+04& .1801+05& .4514+04& .7952+04&-.9647+04&\\
$c_4$& .5803+03& .1804+05&-.4351+05&-.7014+04&-.2607+03&-.2251+05&\\
$c_5$&-.9110+03&-.2210+05& .4724+05& .8324+03&-.1811+05& .5188+05&\\
$c_6$& .5252+03& .1020+05&-.1874+05& .2882+04& .1317+05&-.2717+05&\\
&&&&&&&\\
$\omega^2$&.1823+01&.4865+01&.5895+01&.9113+01&.1465+02&.4228+02&.3226+03\\
&&&&&&&\\
$c_1$& .3028+02&-.7086+02& .1529+03&-.3129+02& .1561+03&-.4624+02& .2042+04\\
$c_2$& .4812+02& .6908+03&-.2810+04& .1313+04&-.1513+04&-.2762+04&-.1461+05\\
$c_3$&-.1305+03&-.3993+04& .1702+05&-.5686+04& .6685+04& .1077+05& .2271+05\\
$c_4$& .2576+03& .8181+04&-.4788+05& .3425+04&-.3673+04& .1875+04& .4154+05\\
$c_5$& .1303+03&-.3086+04& .6823+05& .2433+05&-.2910+05&-.4718+05&-.1496+06\\
$c_6$&-.7534+03&-.7924+04&-.4771+05&-.4855+05& .5132+05& .5873+05& .1425+06\\
$c_7$& .5475+03& .6707+04& .1302+05& .2568+05&-.2386+05&-.2120+05&-.4423+05\\
\hline
&$pn_1$&$pn_2$&$pn_3$&$pn_4$&$pn_5$&$pn_6$&$pn_7$
\end{tabular}
\end{center}
\newpage
\begin{center}
Table 2.\vspace{.2cm}\\
\begin{tabular}{crrrrrrr}\hline
$\omega^2$&.1534+01& .4836+01& .9473+01& .1938+02& .4083+02& .1128+03&\\
&&&&&&&\\
$c_1$& .9752+02&-.6975+02& .2464+03&-.2246+03&-.9102+03& .3169+04&\\
$c_2$& .3284+02&-.8725+03&-.1121+04&-.2590+04& .1713+05&-.2631+05& \\
$c_3$& .2096+03& .3859+04& .5591+04& .1444+05&-.1023+06& .6390+05& \\
$c_4$&-.5354+03&-.5728+04&-.1216+05&-.9903+04& .2599+06&-.3406+05& \\
$c_5$& .3941+03& .2528+04& .5215+04&-.2221+05&-.2933+06&-.4814+05& \\
$c_6$& .1803+01& .1125+04& .3307+04& .2153+05& .1208+06& .4268+05& \\
&&&&&&&\\
$\omega^2$&.1533+01& .4688+01& .7993+01& .9068+01& .1124+02& .1909+02& .1093+03\\
&&&&&&&\\
$c_1$& .9318+02&-.1440+03&-.1202+03&-.1069+04&-.5706+03&-.5482+02& .3703+04\\
$c_2$& .1121+03& .6997+03& .5482+04& .1856+05& .7685+04&-.5626+04&-.3381+05\\
$c_3$&-.2118+03&-.4506+04&-.2955+05&-.1063+06&-.4112+05& .3078+05& .1007+06\\
$c_4$& .2709+03& .9777+04& .5298+05& .2726+06& .7791+05&-.4371+05&-.1109+06\\
$c_5$& .1206+03&-.9309+03&-.6283+04&-.3375+06&-.1278+05&-.7049+03& .1239+05\\
$c_6$&-.7005+03&-.1574+05&-.7154+05& .1894+06&-.9027+05& .3228+05& .4581+05\\
$c_7$& .5309+03& .1200+05& .5087+05&-.3511+05& .5945+05&-.1218+05&-.1722+05\\
\hline
&$pn_1$&$pn_2$&$pn_3$&$pn_4$&$pn_5$&$pn_6$&$pn_7$
\end{tabular}
\end{center}
\newpage
\begin{center}
Table 3.\vspace{.2cm}\\
\begin{tabular}{crrrrrrr}\hline
$\omega^2$&.7569+00& .2822+01& .5661+01& .8814+01& .1519+02& .6952+02&\\
&&&&&&&\\
$c_1$& .6291+03&-.1067+04& .2143+04&-.1949+04&-.6870+04& .1400+05&\\
$c_2$&-.9217+02& .1770+04&-.1693+05& .1131+05& .8373+05&-.2337+06&\\
$c_3$& .4162+03& .2808+04& .5682+05&-.3654+04&-.3195+06& .1293+07&\\
$c_4$&-.3883+04& .5860+04&-.1184+06&-.2807+05& .4791+06&-.3112+07&\\
$c_5$& .6427+04&-.2303+05& .1257+06&-.4668+04&-.2545+06& .3371+07&\\
$c_6$&-.3089+04& .1612+05&-.4514+05& .3416+05& .1251+05&-.1344+07&\\
&&&&&&&\\
$\omega^2$&.7569+00& .2813+01& .5021+01& .8747+01& .1272+02& .3322+02& .7683+02\\
&&&&&&&\\
$c_1$& .5590+03&-.8716+03& .2653+03&-.2421+04& .1881+04& .1412+05& .3376+05\\
$c_2$& .1189+04&-.2018+04& .1406+05& .1926+05&-.7436+04&-.2356+06&-.5191+06\\
$c_3$&-.6377+04& .2349+05&-.1057+06&-.4732+05&-.5363+05& .1298+07& .2528+07\\
$c_4$& .9376+04&-.3509+05& .2059+06& .6165+05& .2228+06&-.3112+07&-.4750+07\\
$c_5$& .5449+03&-.4645+04&-.2977+05&-.4272+05&-.7106+05& .3356+07& .2298+07\\
$c_6$&-.1192+05& .4364+05&-.2533+06&-.3854+05&-.4046+06&-.1333+07& .2455+07\\
$c_7$& .7228+04&-.2275+05& .1775+06& .5845+05& .3227+06&-.1382+03&-.2085+07\\
\hline
&$pn_1$&$pn_2$&$pn_3$&$pn_4$&$pn_5$&$pn_6$&$pn_7$
\end{tabular}
\end{center}
\newpage
\begin{center}
Table 4.\vspace{.2cm}\\
\begin{tabular}{crrrrrrr}\hline
$\omega^2$&.4481+00& .1827+01& .4078+01& .6515+01& .1170+02& .1391+03&\\
&&&&&&&\\
$c_1$&-.2888+02& .1663+03&-.2794+03& .1593+03& .1405+03& .1081+05&\\
$c_2$&-.2440+03&-.7593+04& .2050+05&-.2099+05& .2665+05&-.2129+06&\\
$c_3$& .4933+05&-.2772+04&-.1400+06& .1883+06&-.3467+06& .1344+07&\\
$c_4$&-.1722+06& .1443+06& .2902+06&-.5138+06& .1372+07&-.3583+07&\\
$c_5$& .2124+06&-.2675+06&-.2194+06& .4871+06&-.2092+07& .4207+07&\\
$c_6$&-.8916+05& .1394+06& .5712+05&-.1179+06& .1073+07&-.1790+07&\\
&&&&&&&\\
$\omega^2$&.4380+00& .1805+01& .4006+01& .6190+01& .7980+01& .1439+02& .8964+02\\
&&&&&&&\\
$c_1$&-.1701+02& .1379+03&-.3341+03& .3427+03&-.3020+03& .7695+03& .8642+04\\
$c_2$&-.6649+03&-.6322+04& .2326+05&-.3097+05& .2196+05&-.1111+05&-.1534+06\\
$c_3$& .5135+05&-.1143+05&-.1601+06& .2940+06&-.2552+06& .1349+06& .8174+06\\
$c_4$&-.1667+06& .1599+06& .3264+06&-.9227+06& .1022+07&-.9712+06&-.1565+07\\
$c_5$& .1694+06&-.2551+06&-.1784+06& .1132+07&-.1574+07& .3018+07& .4968+06\\
$c_6$&-.1770+05& .8656+05&-.9582+05&-.4879+06& .7432+06&-.3959+07& .1421+07\\
$c_7$&-.3646+05& .3341+05& .9586+05& .2938+05& .8318+05& .1819+07&-.1047+07\\
\hline
&$pn_1$&$pn_2$&$pn_3$&$pn_4$&$pn_5$&$pn_6$&$pn_7$
\end{tabular}
\end{center}
\end{document}